# Great expectations in music: violation of rhythmic expectancies elicits late frontal gamma activity nested in theta oscillations

## Late elicited gamma activity

Edalati, M.[1,2], Mahmoudzadeh, M.[1,3], Safaie, J.[2], Wallois, F.[1,3*], and Moghimi, S.[1,2,4*]

[1] Inserm UMR1105, Groupe de Recherches sur l'Analyse Multimodale de la Fonction Cérébrale, CURS, Avenue Laennec, 80036 Amiens Cedex

[2] Electrical Engineering Department, Ferdowsi University of Mashhad, 9177948974 Mashhad, Iran.

[3] Inserm UMR1105, EFSN Pédiatriques, CHU Amiens sud, Avenue Laennec, 80054 Amiens Cedex

[4] Rayan Center for Neuroscience and Behavior, Ferdowsi University of Mashhad, 9177948974 Mashhad, Iran.

* Corresponding authors: Fabrice Wallois, Sahar Moghimi

fabrice.wallois@u-picardie.fr

sahar.moghimi@inserm.fr

# Great expectations in music: violation of rhythmic expectancies elicits late frontal gamma activity nested in theta oscillations

**Abstract-** Rhythm processing involves building expectations according to the hierarchical temporal structure of auditory events. Although rhythm processing has been addressed in the context of predictive coding, the properties of the oscillatory response in different cortical areas is still not clear. We explored the oscillatory properties of the neural response to rhythmic incongruence and explored the cross-frequency coupling between multiple frequencies to provide links between the concepts of predictive coding and rhythm perception. We designed an experiment to investigate the neural response to rhythmic deviations in which the tone either arrived earlier than expected or the tone in the same metrical position was omitted. These two manipulations modulate the rhythmic structure differently, with the former creating a larger violation of the general structure of the musical stimulus than the latter. Both deviations resulted in an MMN response, whereas only the rhythmic deviant resulted in a subsequent P3a. Rhythmic deviants due to the early occurrence of a tone, but not omission deviants, elicited a late high gamma response (60-80 Hz) at the end of the P3a over the left frontal region, which, interestingly, correlated with the P3a amplitude over the same region and was also nested in theta oscillations. The timing of the elicited high-frequency gamma oscillations related to rhythmic deviation suggests that it might be related to the update of the predictive neural model, corresponding to the temporal structure of the events in higher-level cortical areas.

**Keywords: Predictive coding, prediction error, event related potential, mismatch negativity, phase amplitude coupling.**

**Introduction**

Music consists of organized sequences of sounds, arranged in hierarchical temporal patterns that unfold over time and involve complex cognitive processes (Patel and Daniele, 2003; Pearce and Wiggins, 2012). Music perception involves the generation of expectations, anticipation of their development, and eventually violation or fulfilment of the predictions (Cheung et al., 2019; Friston, 2002, 2010; Gold et al., 2019; Lumaca et al., 2018). Generally speaking, music plays with our expectations in two forms; it plays with *what* to expect and *when* to expect an event. Although the part of *what* to expect is shaped through melody, phrase, and harmonic structures, the part of *when* to expect an event involves matching the rhythmic structure of music with rhythmic or metrical templates that can be extrapolated in the future (Rohrmeier and Koelsch, 2012).

Recently, music perception has been addressed in the context of predictive coding (Cheung et al., 2019; Friston, 2002, 2005; Gold et al., 2019; Koelsch et al., 2019; Rohrmeier and Koelsch, 2012), which provides a compelling framework to address the predictive processes. This view is based on hierarchical Bayesian inference (Friston, 2005, 2010) and assumes that the brain constantly models statistical regularities in the auditory stream, actively produces predictions that are compared with incoming auditory inputs, and optimizes representations to reduce prediction error (Koelsch et al., 2019). It has been suggested that when exposed to music, the listener's brain extracts the temporal regularities, as well as the rhythmic structures, and shapes a probabilistic model, which in turn provides predictions on *when* to expect an event (Koelsch et al., 2019; Lumaca et al., 2018; Vuust et al., 2009). If the auditory input violates the prediction of the neural model, e.g. if a certain note arrives earlier/later than expected, an *error* signal is generated - the more accurate the prediction, the smaller the prediction error (Hansen and Pearce, 2014). The brain is functionally organized to minimize this *error* to accomplish temporally precise predictions. This attempt is shaped by a reciprocal cascade of cortical functions, in which higher-level structures generate predictions of inputs from lower-level ones and pass them through top-down connections. Then, error signals are transferred through bottom-up connections to update the models that led to these predictions (Kanai et al., 2015).

The neural response to violations of rhythmic structures has been addressed mostly in the context of specific auditory evoked potentials, the mismatch negativity (MMN) (Bouwer and Honing, 2015; Bouwer et al., 2014; Geiser et al., 2009; Grahn, 2012; Honing et al., 2009; Lappe et al.,

2016; Lappe et al., 2013; Lelo-de-Larrea-Mancera et al., 2017; Vuust et al., 2016; Zhao et al., 2017), as well as later ERP components, including P3 (Friedman et al., 2001). Rhythmic deviations, in terms of a tone occurring earlier than expected, elicit an MMN between 100-200 ms (Lumaca et al., 2018), which, depending on the experimental design, can be followed by a subsequent P3a in a time window of ~200-300 ms (Geiser et al., 2009; Vuust et al., 2016; Vuust et al., 2009). In addition, the pre-attentive neural response to the occasional omission of tones within a rhythmic sequence manifests as an MMN response (Bouwer et al., 2014; Honing et al., 2009; Ladinig et al., 2009), which is followed by a P3a component in some paradigm designs (Bouwer et al., 2016). The theory of predictive coding has been used to explain the neural response to rhythmic incongruence, with the MMN having the properties of an error term and the P3a reflecting the subsequent evaluation (Vuust et al., 2009). In this view, the relatively less pronounced MMN response to small rhythmic violations in complex rhythmic patterns has been related to less confident predictions of the weaker neural models - the more difficult the stimuli are to model, the weaker the predictive models, and hence the smaller the prediction error (Lumaca et al., 2018). Consistent with this view, it has also been demonstrated that the amplitude of the MMN depends on the metrical position of the omitted tone, being stronger for metrically stronger positions than metrically weaker ones (Bouwer et al., 2014; Bouwer et al., 2016; Ladinig et al., 2009), again eliciting a more pronounced MMN in response to larger violations. Furthermore, the amplitude of the MMN is modulated by musical training (Lappe et al., 2013; Lappe et al., 2011) and is stronger in musicians than non-musicians (James et al., 2012; Vuust et al., 2009; Vuust et al., 2005), probably reflecting a stronger metrical predictive structure in experienced listeners. In addition, it has been suggested that musical cultural backgrounds shape expectations toward rhythmic structures implicitly through music exposure throughout life, which in turn can modulate the error signal created in response to rhythmic incongruence (Akrami and Moghimi, 2017; Haumann et al., 2018). Put together, the statistical regularities and hierarchical nature of rhythmic structures make music rhythm a powerful tool to investigate predictive coding in the brain, which in turn can be employed to explain the neural dynamics underlying the perception of rhythm.

The mechanisms underlying the reciprocal relationships between predictions and prediction errors have been investigated using several experimental paradigms that rely on the contrast between neural responses to anticipated and novel auditory stimuli (Garrido et al., 2008; Garrido et al., 2007). An elegant paradigm, referred to as the ''local-global'' paradigm, has been developed and used to address the auditory novelty response, as well as dissociate predictions based on local probabilities from those related to global rules (Chao et al., 2018; Chennu et al., 2013). It has been demonstrated that the detection of the violation of the global rules, in which subjects have to create a neural model of the temporal pattern and non-local dependencies of tones, results in a more global and integrative violation of expectation, which manifests as both early and late ERP components (Chennu et al., 2016). Violations of the global rule of stimuli sequence also elicit widespread and protracted oscillatory responses (Dürschmid et al., 2016), including low-frequency theta/alpha effects (Recasens et al., 2018), fronto-temporo-parietal depression in the beta-band (Dürschmid et al., 2016), and high gamma augmentation in the temporal and frontal areas (El Karoui et al., 2014; Kaiser et al., 2005; Kaiser et al., 2007). The different oscillatory activities over distinct brain regions, as well as early/late ERP components during the processing of low-level violations versus violations of the global rules, reflect the different underlying neural mechanisms recruited for the aforementioned processes. In the context of this paradigm, the response to a deviant tone has also been compared to that of the omission of an expected tone (Wacongne et al., 2011). It has been demonstrated that the relatively late ERP component (200-300 ms) present in the processing of mismatch (in which the stimulus differs from that predicted by the recent history of the stimuli) is not elicited during the processing of omission (lack of any sensory input) (Chennu et al., 2016).

Generally, the predictive coding framework suggests that lower-level violations arise from the primary auditory cortex, whereas violations of the global rule of sequences, which require revising the mental representation of the sequence in the higher-level system, involve (1) the prefrontal cortex (Bekinschtein et al., 2009; Chao et al., 2018; Chennu et al., 2013; El Karoui et al., 2014; Uhrig et al., 2014) and (2) updating the predictions for the next trial in lower-level sensory areas (Chao et al., 2018). Deviant auditory stimuli evoke neural responses in the bilateral auditory cortex, superior temporal gyri, and prefrontal cortex (Doeller et al., 2003; Molholm et al., 2005; Rinne et al., 2005). Beyond the auditory cortex, the prefrontal cortices integrate error signals to update the prediction models (Bastos et al., 2012; Summerfield et al., 2006). Indeed, there is evidence for a frontotemporal hierarchy of prediction and prediction error information transfer (Chao et al., 2018; Chennu et al., 2016; Garrido et al., 2008; Garrido et al., 2007, 2009; Phillips et al., 2015).

Music rhythmic violations are a good example of the violation of the global rule of sequences, in which the expectation is developed based on the modeled local and non-local temporal dependencies. We designed an experiment to investigate the neural response to rhythmic deviations in which the tone either arrived earlier than expected or the tone in the same metrical position was omitted. These two manipulations modulate the rhythmic structure differently, with the former creating a larger violation of the general structure of the musical stimulus than the latter, in which the Gestalt characteristics of the chord sequence are not violated. Although the processing of music rhythm has been addressed in the context of predictive coding, the properties of the oscillatory response in different cortical areas is still not clear. To date, mostly ERP components have been analyzed to address the neural mechanisms of rhythm perception. To better understand the mechanisms underlying the neural response to rhythmic incongruence and provide the links between the concepts of predictive coding and the perception of rhythm, we explored the oscillatory properties of these responses and compared them under the two conditions of manipulation. In addition, growing evidence suggests that perception involves cross-frequency coupling (CFC) in terms of coordinated slow and fast neural oscillations, typically nested theta/gamma oscillations (Buzsáki and Draguhn, 2004; Canolty et al., 2006; Lakatos et al., 2005), which presumably enhance combinatorial opportunities for encoding (Rasch and Born, 2013) and facilitate synaptic plasticity (Bergmann and Born, 2018; Buzsáki and Draguhn, 2004; Salimpour and Anderson, 2019). Thus, given the role of CFC in perception, we explored the CFC between multiple frequencies to explain the underlying mechanisms involved in rhythm perception. Our central hypothesis was that there should be a fundamental difference in the neural response to violations consisting of the omission of tones and rhythmic violations due to tones arriving earlier than expected, with the latter creating a larger violation of the rhythmic structure. We hypothesized that the greater violation of the rhythmic structure would elicit stronger late oscillatory activities following the MMN (which in turn reflects the prediction error), which would be related to the update of the neural model of the rhythmic structure. In agreement with our hypothesis, both deviations resulted in an MMN response, whereas only the latter resulted in a subsequent P3a. Rhythmic deviation due to the early occurrence of a tone elicited a late high gamma response (60-80 Hz) at the end of the P3a over the left frontal region, which interestingly correlated with the P3a amplitude over the same region and was also nested in theta oscillations. The timing of the elicited high-frequency gamma oscillations, which is probably after the integration of the bottom-up and top-down processing, suggests that might be related to the update of the predictive neural model corresponding to the temporal structure of the events in higher-level cortical areas.

## Materials and methods

### Participants

Fourteen right-handed healthy volunteers (age 20 ± 2 years, 7 females) participated in the study after providing written informed consent. Participants were non-musicians with a similar educational background (undergraduates or MSc students), normal hearing, and normal or corrected-to-normal vision. They reported normal nocturnal sleep patterns (7–9 h starting between 10 pm and 12 am) for the week before the experiment. They had not used caffeine, nicotine, or energy drinks on the day of the experiment and had not performed excessive exercise within the previous 24 h. As assessed by a questionnaire, participants had no history of neurological or psychiatric disorders.

### Auditory stimuli and the experimental paradigm

The stimulus material consisted of an auditory rhythm in 2/4 meter on one chord that was presented repeatedly and continuously at 100 bpm with a piano sound. The standard rhythm consisted of a chord with a duration of a quarter note at the beginning of each bar followed by two eighth notes. The rhythmic changes (rhythm deviant) consisted of replacing the two eighth notes with one sixteenth note, one eighth note, and one sixteenth rest. The omission deviant was created by removing the last chord. Analyzing a rhythmic sequence, one can imagine a "tree" structure, corresponding to the hierarchical representation of a sequence of timed events (notes in music). In this tree, the "root" node at the highest level of the hierarchy is considered as the whole bar, the nonterminal nodes signify the lower level metrical units, and the terminal nodes of the tree are all either (sounded) notes or rests (Longuet-Higgins and Lee, 1984). Three rhythm trees are presented in Fig S1 showing the rhythmic structure of the standard stimulus and the rhythm and omission deviants. This figure shows how the rhythm deviant induces a new branch in the tree.

We distracted the participants from the main objective of the experiment by adding two other deviant conditions. The frequency of the last note was changed to create a pitch deviant. Finally, we included a timbre deviant, for which one chord during the bar (randomly defined) was played by a violin sound. Stimuli were constructed using open-source MuseScore 2 software and exported as wav-files. A dynamic accent of 25 percent above the general intensity was induced on the first beat of each stimulus to reinforce the perceived meter (Geiser et al., 2009). This accent is indicated by a ''>'' in Fig 1B, in which the stimuli and experimental protocol are depicted.

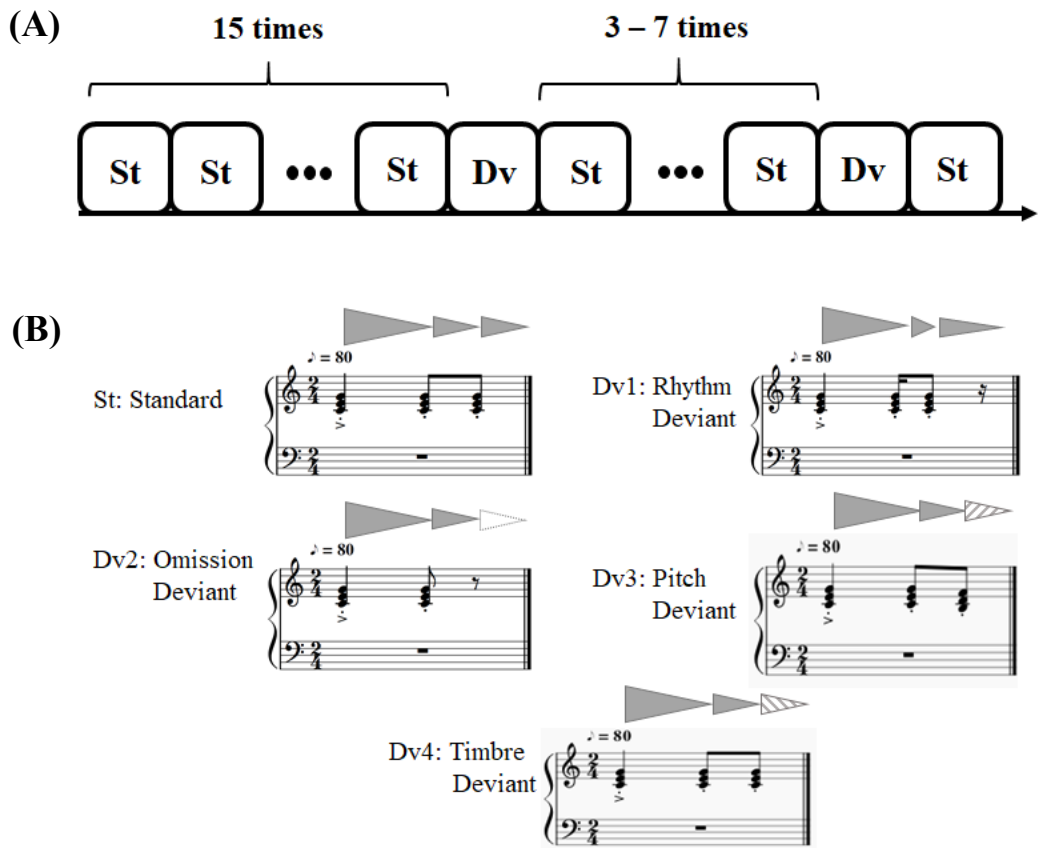

Figure 1. (A) The Experimental protocol. (B) The standard and deviant stimuli.

The stimuli were delivered in the context of an oddball paradigm. The experimental session included high-probability standard stimuli (p = 0.85, 2,850 trials) interspersed with four infrequent deviant patterns, accounting for the remaining stimuli (p = 0.15, 150 trials of the timbre deviant, and 105 trials for the rhythm, omission, and pitch deviants each). The order of the four deviant stimuli was pseudorandomized among the standard trials, enforcing three to seven standard stimuli between successive deviant trials. Control stimulus blocks, corresponding to rhythm and omission deviants, presented 200 times continuously without any standard trial in between, were used to evaluate the response to the rhythm and omission deviant trials in the main experiment. These blocks were presented randomly in the main protocol.

We then performed another control experiment during which the rhythm deviant stimulus replaced the standard condition and the deviant condition consisted of silencing its last chord in the context of an oddball paradigm. This control experiment allowed us to further explain the variation in the neural response due to the omission of the tone from that observed in the rhythm deviant condition (see SI for more details).

Stimuli were delivered through two custom-made speakers at 65 dB SPL using Psychtoolbox MATLAB. Participants sat on a comfortable chair in dim light and were instructed to watch a silent movie (March of the Penguins, Warner, ASIN B000BI5KV0) and to press a button whenever they detected a timbre deviant. The total duration of the experiment was ~78 min. The neural response to the timbre deviants, as well as the pitch deviants, were not analysed, since they were included to distract the subject from the main objective of the experimental design.

**EEG acquisition and preprocessing**

High-density EEG Data were acquired using a high impedance amplifier Net Amp 300 and Net Station 5 with a sensor net consisting of 128 electrodes (Geodesic Sensor Net, Electrical Geodesics, Inc., USA). Impedances were kept below 50 kΩ. The EEG was digitized at a 1,000-Hz sampling rate, with a Cz vertex electrode as reference. The recorded signals were analyzed with MATLAB® software (The MathWorks, Inc., Natick, Massachusetts, United States), using FieldTrip (Oostenveld et al., 2011), EEGLAB (Delorme and Makeig, 2004) and custom MATLAB functions and codes. A two-pass 0.5-100 Hz finite impulse response (FIR) bandpass filter (order = 3 cycles of the low frequency cut-off) from the EEGLAB toolbox was applied to remove low and high-frequency artifacts from the EEG signals. Artifact-ridden channels were removed and interpolated. Also, a 50-Hz notch filter was applied to remove the line noise. Artifacts (e.g., eye-blink, eye-movement, and muscle activity) were then removed by independent component analysis (ICA) using the EEGLAB toolbox. The preprocessed data were later epoched starting 650 ms before the onset of the deviant note and ending 800 ms after. Epochs were excluded if the standard deviation of amplitude exceeded 25 μV within two moving windows of 200 and 800 ms or any sampling point exceeded 75 μV at any electrode location. EEG data were later re-referenced to the average reference. After artifact rejection, the number of remaining trials was 87.98 ± 13.47, 176.58 ± 18.24, 87.15 ± 14.78, and 167.4 ± 23.93 for the rhythm deviant, rhythm control, omission deviant, and omission control conditions, respectively.

**Event-related potential (ERP)**

A 25-Hz low-pass FIR filter (13 cycles) was applied to calculate the ERP response. For each trial, zero was set at the onset of the third chord (the expected location of the third chord for the omission condition), with the baseline being set at -500 ms to -300 ms and -650 ms to -450 ms for the rhythm and omission deviants, respectively (for both conditions the baseline was set at 250-450 ms from the onset of the first chord, which was after the disappearance of the response to the first chord). Event-related potentials were computed by averaging the EEG trace of the remaining trials for each condition after baseline correction. A nonparametric cluster-based permutation procedure (1,000 permutations), implemented in the FieldTrip toolbox (Maris and Oostenveld, 2007), was applied to search for significant changes in the deviant condition relative to the control condition. The initial threshold for cluster definition and the minimum number of neighbors were set to $P < 0.05$ and four, respectively. Finally, the final significance threshold for summed t values within clusters was set to $P < 0.05$.

**Time-frequency representation (TFR)**

TFRs were calculated per event epoch ('mtmconvol' function of the FieldTrip toolbox) for frequencies from 4 to 60 Hz (using Hanning tapers) and from 30 to 100 Hz (using discrete prolate spheroidal sequence tapers) in steps of 0.25 Hz. The TFR was calculated using a sliding window with a variable frequency-dependent length that always comprised a full number of cycles (at least two cycles and an at least 100-ms window length). Time-locked TFRs of all epochs were then baseline corrected and averaged per participant. Statistical analysis was used to check for significant power changes corresponding to the deviant condition relative to the control condition. The cluster-based permutation procedure (1,000 permutations), implemented in the FieldTrip toolbox, was applied to correct for multiple comparisons. The initial threshold for cluster definition and the minimum number of neighbors were set to $P < 0.05$ and four, respectively. Finally, the final significance threshold for summed t values within clusters was set to $P < 0.05$.

**Phase amplitude coupling (PAC)**

We applied a method introduced by Tort et al. (Tort et al., 2010) to simultaneously assess PAC for a large number of frequency pairs. For a given frequency pair, extracted 3.6-second epochs (including three trials, the target trial, and the preceding and following trials) – sufficiently long to prevent any edge effects during filtering – were filtered in both frequency ranges. Lower frequencies ranged from 6 to 15 Hz (0.25-Hz increments, the bandwidth was gradually increased from 0.75 Hz to 1.875 Hz) and higher frequencies ranged from 60 to 100 Hz (0.25-Hz increments, the bandwidth was gradually increased from 7.5 Hz to 12.5 Hz). The time series of the lower frequency phase and the higher frequency amplitude were then extracted using the Hilbert transform. The deviant responses (the third chord for the rhythm deviant, and the expected interval of the third chord for the omission condition) were concatenated and the lower-frequency phases binned into eighteen 20° bins spanning the [-π, π] interval and the corresponding mean amplitude of the higher frequency was computed for each phase bin and then normalized by dividing it by the sum over all bins. Next, the deviation of the PAC profile from a uniform distribution was quantified by defining the modulation index (MI) in terms of the Kullback-Lieber distance between the amplitude distribution $P$ and a uniform distribution $U$, $D_{KL}(P,U) = \log(nbins) - H(P)$, where the Shannon entropy $H$ of the distribution $P$ is $H(P) = -\sum_{bin=1}^{N} P(bin) \times \log[P(bin)]$. Briefly, the MI of Tort et al. (Tort et al., 2010) specifically measures deviations from a uniform distribution; if the high-frequency EEG mean amplitude shows no systematic relationship with the low-frequency phase, the high-frequency amplitude in each low-frequency phase bin will tend toward the overall average high-frequency amplitude, resulting in a flat or uniform distribution. The MI ranges from 0 to 1; a value of 0 shows that the mean amplitude is uniformly distributed over the phases and an MI of 1 shows that the mean amplitude has a Dirac-like distribution.

$$MI = \frac{D_{KL}(P,U)}{\log(nbins)}$$

For statistical analysis, the cluster-based permutation procedure, implemented in the FieldTrip toolbox, was used to compare the deviant comodulogram with that corresponding to the control data at the group level. The initial threshold for cluster definition and the minimum number of neighboring were set to $P < 0.05$ and four, respectively. The final threshold for significance of the summed $t$ value within clusters was set to $P < 0.05$. In addition, surrogate chance level PAC data at each electrode and frequency band were created by calculating the MI after permutation of the epoch numbers corresponding to amplitude and phase time series for both deviant and control conditions. We obtained a distribution of surrogate MI values for each subject, EEG electrode, and frequency band, by creating 500 surrogate data sets and computing the associated MIs. Finally, the cluster-based permutation procedure was used to compare the empirical MI with that corresponding to the surrogate data at the group level for both the rhythm and omission conditions. The initial threshold for cluster definition and the final threshold for significance of the summed $t$ value within clusters were both set to $P < 0.01$.

We further investigated the nesting of theta-gamma activity (6.5-8.5 Hz for the phase frequency and 61-78 Hz for the amplitude-frequency, the choice of the phase/amplitude frequencies was made based on the results of the comodulogram analysis) in the time course of the deviant response by calculating the PAC over a sliding window of 200 ms in steps of 5 ms. We thus concatenated the 200-ms windows corresponding to each trial (the epochs were cut after applying the filter and Hilbert transform) and performed the aforementioned method to calculate the MI. This procedure was repeated at each time step, which resulted in a MI time-series with a resolution of 5 ms. For statistical analysis, the cluster-based permutation procedure was implemented.

## Results

Our results, presented in detail below, show that both small and large rhythmic deviations induced a neural response with a time course and topographical distribution typical of MMNs. They further demonstrate a significant difference in the amplitude of the MMN between the two conditions. Interestingly, we only observed a P3a for the rhythm deviants, which proceeded the MMN and was concurrent with the emergence of gamma activity over the left frontal area.

### Event-related potentials

The ERP response to the rhythm and omission deviants are depicted in Fig. 2A and B. We considered the baseline time window [-500 ms to -300 ms] for the rhythm and [-650 ms to -450 ms] for the omission condition (that is [250-450 ms] from the onset of the first chord in both conditions). The grand average ERPs showed enhanced early (~100-200 ms) frontal negativity, consistent with the typical time window of MMN, for both deviant conditions with respect to the control conditions (Fig. 2A and B). The MMN amplitude, corresponding to the rhythm condition, was significantly larger than that of the omission condition, as shown by a paired samples t-test (t = 3.258, p = 0.0076). There was no significant difference between the latency of MMN corresponding to the two conditions. For the rhythm condition only, the MMN was followed by a subsequent positive deflection in the 200 to 300-ms time window, indicative of a P3a (Fig. 2A). Through visual inspection, both components were more pronounced over the frontal and frontocentral electrodes and demonstrated an inverting polarity over the posterior electrodes.

Cluster-based statistics revealed four spatiotemporal clusters for the rhythm condition (their time intervals are specified by the thick black lines in Fig. 2A): (1) a negative cluster ($p = 0.012$, corrected) comprising frontal and frontocentral electrodes and extending over 120-202-ms post-final chord, (2) a posterior positive cluster ($p = 0.004$, corrected) synchronous with the first cluster, 114-204-ms post-final chord, (3) a negative posterior cluster ($p = 0.036$, corrected) extending over 221-290-ms post-final chord, and (4) a positive frontal cluster ($p = 0.042$, corrected) synchronous with the third cluster, 217-294-ms post-final chord. Spearman correlation analysis revealed a significant correlation between the MMN amplitude, averaged over 130-180 ms, and the P3a amplitude, averaged over 200-300 ms ($r = 0.8909$, $p = 0.0014$), over the frontocentral cluster. Cluster-based statistics revealed two spatiotemporal clusters for the omission condition (their time intervals are specified by the thick black line in Fig. 2B): (1) a negative cluster ($p = 0.026$, corrected) comprising frontal and frontocentral electrodes and extending over 133-221-ms post-final chord and (2) a posterior positive cluster ($p = 0.0022$, corrected) synchronous with the first cluster, 137-215-ms post-final chord. We also performed a control experiment during which the rhythm deviant stimulus replaced the standard condition in the context of an oddball paradigm and the deviant condition consisted of silencing its last chord (see SI for more details). During this control experiment, omission of the last chord in Dv1 (Fig. S1B) resulted in a smaller MMN than the rhythm deviant response in the main experiment and it was not followed by the significant P3a that was observed in the rhythm deviant condition during the main experiment (Fig S2). The results of this control study confirm that the omission condition results in a relatively smaller MMN and does not elicit a significant P3a component.

The topographical distribution of the clusters corresponding to different time windows in the course of the deviant response are shown in supplementary Fig. S3. In addition, the supplementary movie illustrates the evolution of the ERP response over the scalp during the course of the deviant response for both rhythm and omission conditions.

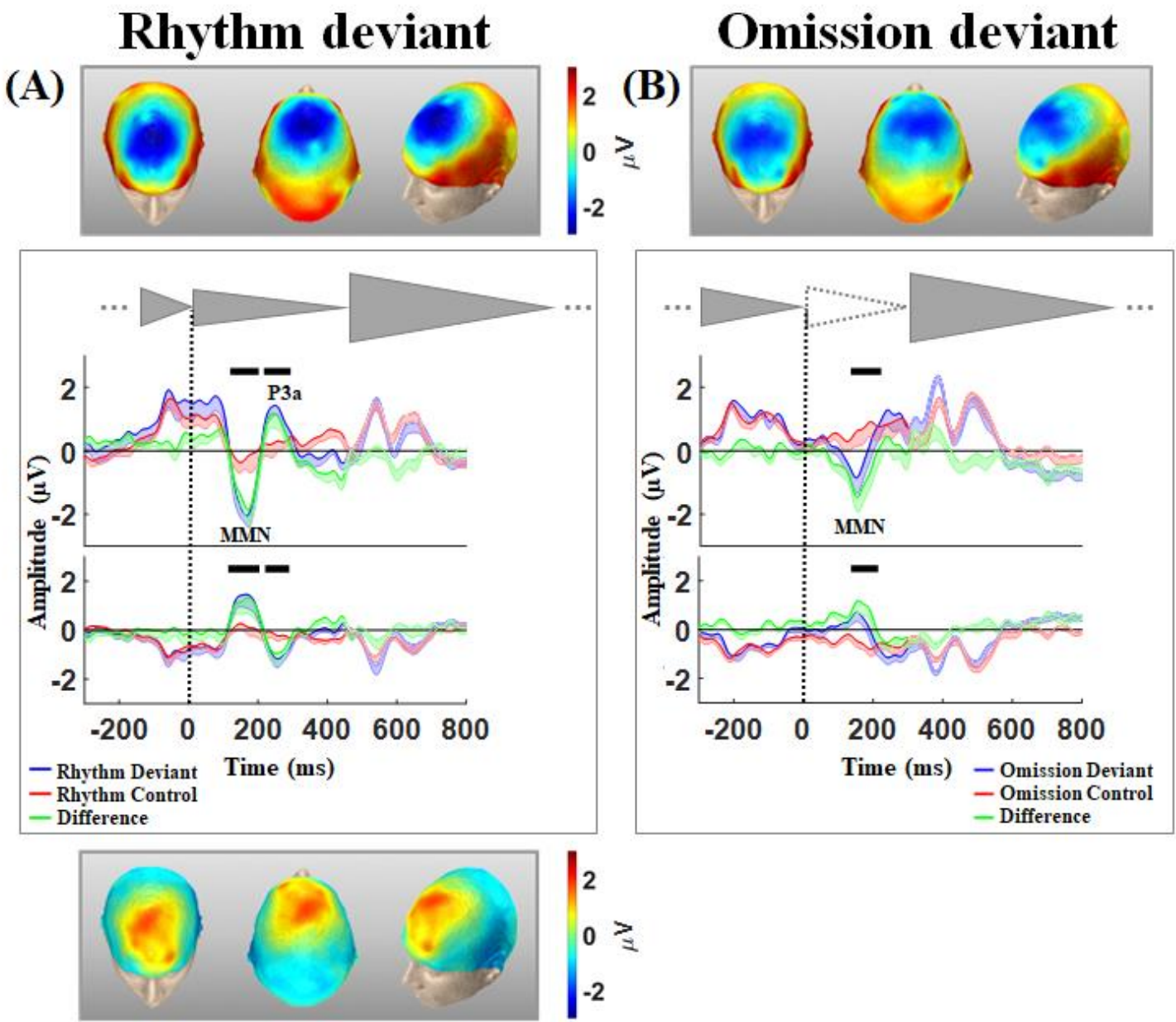

**Figure 2. Event-locked analysis of rhythm and omission conditions.** The onset of the deviant chord was set to zero and the next trial started at 450 ms and 300 ms for the rhythm and omission conditions, respectively. For both conditions, the baseline was set to 250 to 450 ms from the onset of the first chord. (A) Grand average of ERP (+SE) for the rhythm deviant condition, the control condition, and their difference over frontal and parietal clusters. (B) Same as (A) for the omission condition. Both Rhythm and Omission deviants elicited an MMN. However, a P3a followed the MMN for only the rhythm deviants. The black bars over the ERP figures represent the time intervals of significant difference between the deviant and control conditions ($p < 0.05$, corrected, marked according to cluster-based permutation analysis). The topography of each significant time window is shown in the boxes: above for the MMN corresponding to the rhythm and omission deviant conditions and below for the P3a in the rhythm deviant condition.

**Time-frequency representation**

For each participant, we calculated the TFR over 4 to 30 Hz and 60 to 100 Hz and set the zero to the onset of the deviant condition to determine the power modulation during the time course of the deviant response (Fig. 3). The mean TFRs over the left, middle, and right frontal electrodes were computed to show the spatial dynamic of power fluctuations in the low- (4-30 Hz) and high-frequency (60-100 Hz) ranges for the rhythm and omission conditions (deviant minus control), respectively (Fig. 3A and B).

As illustrated, the pronounced oscillatory power over the 4- to 10-Hz frequency range coincided with the temporal location of MMN. Cluster-based statistics revealed a spatiotemporal cluster over 90 to 228 ms ($p = 0.002$, corrected) for the rhythm condition and over 140 to 283 ms ($p = 0.001$, corrected) for the omission condition over the 4- to 10-Hz frequency range, comprising the frontal and frontocentral electrodes (clusters are presented in Fig.3). The difference in the mean power and time of emergence of the low-frequency clusters, corresponding to the rhythm and omission deviant conditions, were not significant. However, on average, the significant low-frequency cluster arrived earlier for the rhythm condition, as shown by visual comparison of Fig. 3A and B, and the amplitude of the TFR was more pronounced for the rhythm than omission condition.

In the rhythm deviant trials, TFR analysis over the 70- to 100-Hz frequency range (Fig. 3A) showed pronounced high-frequency oscillatory power over the left frontal electrodes. The significant difference relative to the control condition was restricted to the 260 to 289-ms time window ($p = 0.019$, corrected) after the onset of the deviant chord (Fig. 3A), concurrent with the descending slope of the P3a component. Interestingly, the high-frequency oscillatory power, as averaged between 200 and 300 ms, correlated with the P3a amplitude over the left frontal cluster ($r = 0.6909$, $p = 0.0231$), as revealed by Spearman correlation analysis.

In the omission deviant trials, TFR analysis over the 60- to 80-Hz frequency range (Fig. 3B) showed a non-significant increase in high-frequency oscillatory power over the right frontal electrodes between 100 and 300 ms, followed by a significant decrease comprising five right frontal electrodes and extending from 307 to 330 ms ($p = 0.0475$, corrected, Fig. 3B), which coincided with the onset of the next standard trial.

**Phase amplitude coupling**

To further address the underlying mechanisms of rhythm perception and investigate the inter-relationships of oscillatory activities during the processing of rhythm deviations, we evaluated PAC across a broader frequency range by applying the comodulogram analysis (Tort et al., 2010). The modulation index (MI) reflects the degree to which the amplitude of the higher (modulated) frequency varies as a function of the phase of the lower (modulating) frequency. We performed the PAC analysis over all electrodes for both deviant conditions, with the phase frequency ranging from 6 to 15 Hz and the amplitude frequency ranging from 60 to 100 Hz, as explained in Materials and Methods. Cluster-based statistics revealed a significant cluster ($p = 0.028$, corrected) when comparing the 450-ms window between the rhythm deviant condition with the control condition. This cluster demonstrated the presence of significant PAC for the rhythm deviant condition, in which the power of the 60- to 75-Hz frequency range was modulated by the phase of the 6.5- to 8.5-Hz frequency range (Fig. 3C). In addition, we assessed the statistical significance of the observed CFC by comparing the results with those generated with the epoch-shuffled surrogate data - the same data as for the original PAC analysis, and with exactly the same spectral power characteristics. This procedure showed a single positive cluster ($p = 0.002$, corrected, Fig. S4). This cluster corresponded to the theta-gamma PAC and was also observed when comparing the rhythm deviant and control conditions, demonstrating that the observed effect was due to a pronounced increase in theta-gamma coupling in the rhythm deviant condition. There was no significant cluster, either when comparing the omission deviant condition with the control condition or when comparing the former with the shuffled surrogate data.

We further investigated the temporal pattern of PAC during the time course of the deviant response. The mean time-varying PAC over subjects for the rhythm deviant condition is illustrated in Fig. 3D. The plotted MI is the time-varying deviant MI minus the time-varying control MI over 6.5 to 8.5 Hz for the phase frequency and 61 to 78 Hz for the amplitude-frequency. The difference between the two conditions was significant and increased over a time window from 170 to 260 ms ($p = 0.0459$, corrected) from the onset of the deviant chord. Interestingly, the timing of the PAC for the deviant condition was concurrent with the late parts of the low frequency TFR cluster and coincided with the significant high-frequency TFR cluster. These PAC results not only corroborate the main findings from our event-based analysis but also highlight that the observed late high-frequency oscillatory activity in the TFR analysis was nested in the low-frequency oscillations in the theta range. There was no significant difference between the PAC corresponding to the omission deviant condition compared to the control condition in the time course of the deviant response.

# Rhythm Deviant

# Omission Deviant

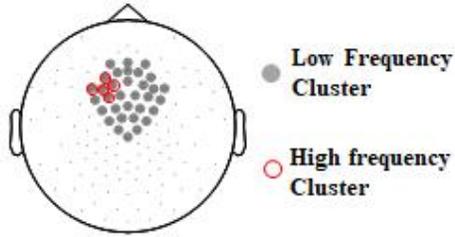
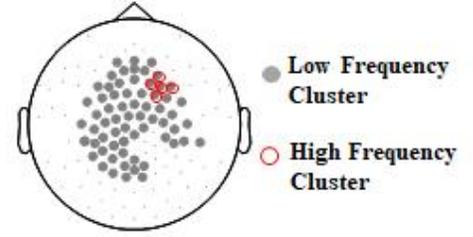
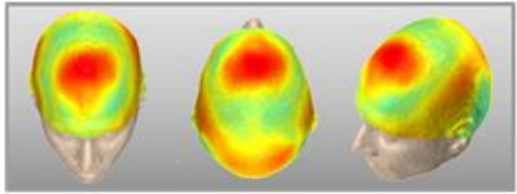
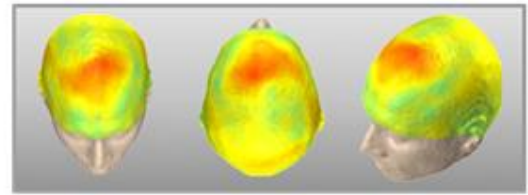
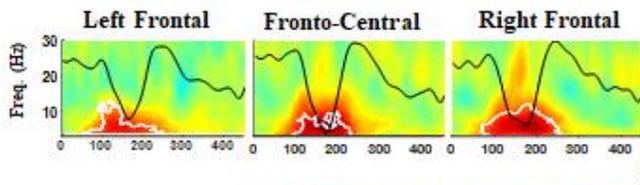
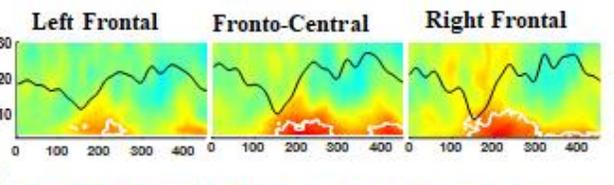
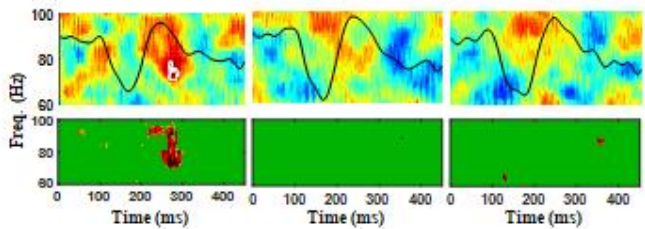
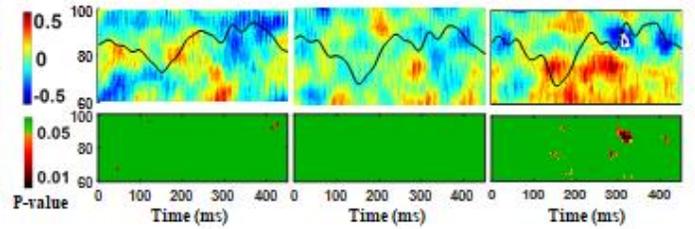
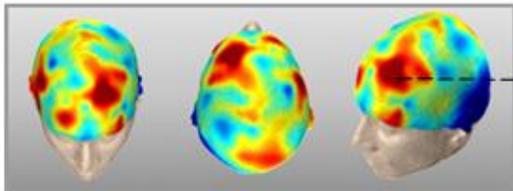
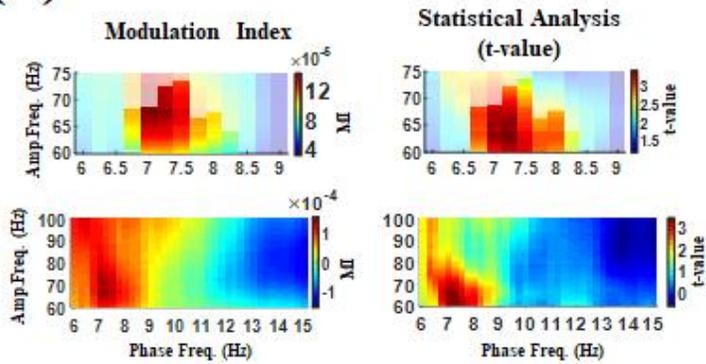
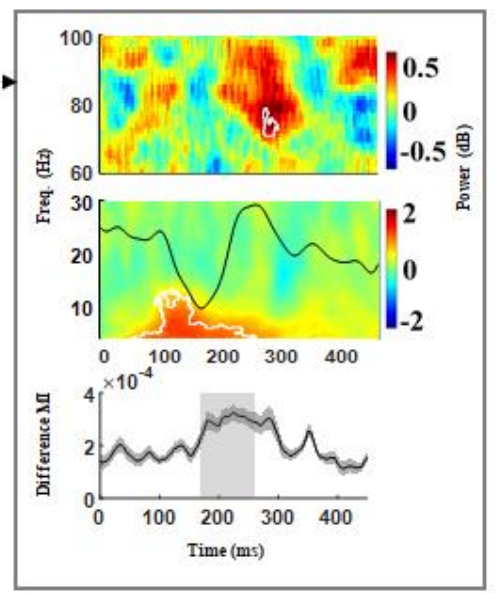

**Figure 3. Event-locked analysis of the rhythm and omission deviant conditions.** The average TFR locked to the beginning of the rhythm deviant (A) and omission deviant (B) trials. The corresponding ERP of the ROI is superimposed on each TFR to better illustrate the results. The statistically significant changes from the control condition are indicated by a white contour. Rhythm deviant: low frequency cluster: 90 to 228 ms, p = 0.002, corrected; high frequency cluster: 260 to 289 ms p = 0.019, corrected. Omission deviant: low frequency cluster: 140 to 283 ms, p=0.001, corrected; high frequency cluster: 307 to 330 ms, p = 0.0475, corrected. The figures below the high frequency TFRs show the uncorrected p values corresponding to the comparison between the deviant and control conditions (paired-sample t-test). The topographical distributions of the electrodes belonging to the significant low- and high-frequency clusters are specified on the head map on top. The topographical distribution of the average power over the frequency and time window corresponding to each cluster is presented in the boxes (above for the low frequency and below for the high frequency TFR). (C) Comodulograms of phase-amplitude coupling analysis over the 450-ms window corresponding to the rhythm deviant condition. Comparison of the rhythm deviant condition with the control condition showed a single cluster with significant MI (p = 0.028, corrected). (D) The difference between the rhythm deviant and control conditions in low-frequency TFR, high-frequency TFR, and time-varying PAC over the time course of the rhythm deviant condition is shown. The significant cluster observed when comparing the two conditions is marked in all the three figures. The time-varying PAC is presented as the mean ± SE.

## Discussion

Both rhythm and omission deviations induced a typical MMN, similar in time course and topographical distribution, with a significantly higher amplitude for rhythm deviations. In addition, a significant P3a was elicited only for the rhythm deviant. Furthermore, rhythm violation through modulation of the rhythmic structure elicited significant late gamma band activity over the left superior frontal area, which occurred concomitantly with the P3a component. This gamma oscillation was nested in theta oscillations, resulting in significant phase-amplitude coupling. The power of the gamma oscillation correlated with the amplitude of the P3a component over the same ROI for the rhythmic deviant condition only. Omission of the last chord in the rhythmic sequence also elicited an MMN, but this component was not followed by a later P3a-positive component in the left frontal area and did not elicit a significant gamma-band response.

Rhythm perception consists of extracting regularities from the sound stream and shaping temporal expectations about the future events. It is considered to be a Bayesian process (Elliott et al., 2014; Koelsch et al., 2019; Lumaca et al., 2018), which fits with the framework of predictive coding (Friston, 2005). Exposure to a repetitive rhythmic sequence generates a higher order neural model of the temporal structure of the stimulus. Generally, predictive coding proposes that the occurrence of an incongruence of the deviant with the higher-order model creates an error signal in the lower levels of the hierarchy that propagates to the higher levels, resulting in an update of the previously created neural model. Here, we presented two types of deviant conditions: rhythm and omission, in which the former condition was created by the last chord arriving earlier than expected and the latter by omission of the last chord. Although we introduced a sensory error in the omission condition, due to the silenced last chord, this did not create a violation of the rhythmic structure, as the Gestalt characteristics of the chord sequence were not violated. Creating the rhythmic deviant condition by having one chord arrive earlier than expected, followed by silence changes the "tree" structure, with the addition of a new branch, which requires updating of the neural model created by repetitive representation of the standard bar. Conversely, omission of the last chord maintains the tree structure and therefore results in a relatively smaller violation than the rhythmic violation. In other words, the omission deviant reflects a prediction error at the sensory level, which is smaller than that created as a result of a rhythm deviant, and which might not require modulation of the predictive model. Studies of the neural correlates of novelty responses show that a significant MMN is elicited by both small and large deviants, whereas a significant P3a is elicited only by large deviants (Friedman et al., 2001). It has been shown that MMN is sensitive to sensory or lower-level violations, reflecting the detection of deviant events, whereas the later responses arrive only when there is a higher-level violation of the regularity of the underlying structure (Chennu et al., 2013; Wacongne et al., 2011). The P3a component is dependent on top-down expectations (Chao et al., 2018; Chennu et al., 2013). It is associated with the evaluation of the deviant events (Lumaca et al., 2018) and reflects updating of the prediction model, which involves a broad frontoparietal network (Wacongne et al., 2011). The P3a has been linked to musical expectancy, being sensitive to large violations of rhythmic (Vuust et al., 2009), metric (Jongsma et al., 2004), melodic (Trainor et al., 2002), and harmonic (Janata, 1995) structure. In addition, P3a is elicited by violation of the general rules (and not the local rules) of the temporal sequence (Bekinschtein et al., 2009; Chennu et al., 2013), suggesting its role in the processing of the overall structure of the stimuli. In our study and during first-level processing, the MMN was sensitive to the prediction error being larger for the rhythm deviant than the omission deviant, for which the rhythmic tree structure was changed. We suggest that violation of the rhythmic tree structure in the rhythm deviant elicited a larger MMN, which moved to higher areas in the chain of auditory processing, leading to integration with the higher areas and then updating of the predictive model of the rhythmic structure in the higher levels of the hierarchy. This hypothesis is supported by the (*i*) occurrence of the P3a component, the amplitude of which correlated with that of the MMN, and (*ii*) by the gamma band activity nested in the P3a component, with its power being correlated with the P3a amplitude.

The observed late induced gamma activity in response to the rhythm deviant condition, in contrast to the control condition, reflects focal synchronized neural activity (in contrast to observed earlier wide-spread effects) at the left frontal electrodes. Gamma-band activity is investigated in processes related to the auditory system and is suggested to reflect attention, anticipation, and expectation (Snyder and Large, 2005; Sokolov et al., 2004; Zanto et al., 2005), as well as the bundling of auditory features into a unitary percept (Bhattacharya et al., 2001). Induced 'late' gamma activity, which typically emerges later than 200 ms, even in concomitance with the P3 component (Başar-Eroglu and Başar, 1991), is suggested to be a signature of processes such as response selection or context updating (Herrmann et al., 2004). During the creation of a phonetic mismatch response, induced gamma activity (84-88 Hz) follows the evoked mismatch response by 130 ms over the left inferior frontal cortex (Kaiser et al., 2002). Focal increased gamma activity (50-90 Hz) has also been observed over the left superior frontal area in response to an acoustic mismatch in the context of an oddball audiovisual paradigm, suggested to reflect higher-order auditory functions following the mismatch response (Kaiser et al., 2005; Kaiser et al., 2003). It has been proposed that the late gamma activity is specifically related to the match between stimulus-related information and top-down factors, as well as the emergence of an object representation (Noesselt et al., 2003; Tallon-Baudry and Bertrand, 1999). We suggest that the observed induced late gamma-band activity in this study reflects the integration of bottom-up and top-down processing towards refining the predictions of the neural model corresponding to the temporal structure of the events in higher-level cortical areas.

Recent studies in humans using ECoG, EEG, MEG, and fMRI have demonstrated that local error signals are restricted to the primary auditory cortex, whereas error signals corresponding to the violation of the global structure propagate to distributed areas in the frontal cortex (Bekinschtein et al., 2009; Chennu et al., 2013; El Karoui et al., 2014; Wacongne et al., 2011). The frontal cortex encodes the global and abstract characteristics of a sequence (Dehaene et al., 2015; Wang et al., 2015). Signals reflecting the update of the neural model are primarily found in the prefrontal cortex and dorsolateral prefrontal cortex, areas important for working memory-related processing (Curtis and D'Esposito, 2003; Gilbert and Kesner, 2006). Chao et al. suggested that these brain structures "generate and hold an internal representation of the entire sequence of stimuli" and therefore can later generate error signals when an unexpected novel sequence is heard (Chao et al., 2018). We suggest that the elicited gamma-band activity over the left frontal cortex reflects the underlying mechanisms involved in updating of the neural model of the entire rhythmic structure over the frontal areas.

In the hierarchical settings of a predictive model, backward connections deliver predictions to lower levels (Park and Friston, 2013), whereas forward connections transfer prediction errors to upper levels (Koelsch et al., 2019; Park and Friston, 2013). In this framework, prediction tuning, or in other words, model updating, occurs through changing synaptic efficacy (Park and Friston, 2013). This is probably reflected by the increased PAC in our results during the neural response to a rhythm deviant, with the lower-frequency theta oscillations synchronizing the synaptic input toward refinement of the predictive model of the temporal pattern of the music structure, which elicits the local high-frequency gamma activity at the time of the P3a component. However, this hypothesis is speculative at present and requires further studies. The gamma-band activity was nested in the late theta activity. Although the low-frequency oscillatory activity was observed for both the omission and rhythm conditions, the elicited nested gamma activity and significant PAC was only observed for the rhythm deviant. Interestingly, in agreement with the observed gamma-band activity, the PAC index became significant only at the time of the emergence of the gamma-band activity and in coalescence with the P3a component. PAC is a potentially useful measure of coupling between neural oscillations on different timescales. The mechanisms underlying PAC have recently received much attention in both experimental and theoretical studies. It has been suggested that PAC supports the encoding, storage, and retrieval of information (Bergmann and Born, 2018; Fell and Axmacher, 2011). In adults, PAC translates as precise temporal relationships between modulating and modulated frequencies in (for example) the thalamocortical and hippocampal networks during sleep (Staresina et al., 2015), in the hippocampus during the operation of multi-item working memory (Axmacher et al., 2010), and in the cortical networks during cognitive functions (Canolty et al., 2006; Chacko et al., 2018; Combrisson et al., 2017). It has been hypothesized that the phase of the slower oscillation generally reflects greater excitability among postsynaptic neurons, which in turn synchronizes the synaptic input (as reflected by an increase in the amplitude of the faster oscillation) (Bergmann and Born, 2018). In adults, the mechanisms underlying PAC lead to specific temporal patterns of coupling across multiple frequencies at both local and distal sites (Bergmann and Born, 2018; Engel et al., 2013; Hashemi et al., 2019; Hyafil et al., 2015). Such coupling can be modulated by structural changes (Helfrich et al., 2018; Salimpour and Anderson, 2019) and task-related network dynamics (suggesting a functional role for the CFC) (Canolty and Knight, 2010; Combrisson et al., 2017; Engel et al., 2013; Haegens et al., 2011; Tort et al., 2009). The precise PAC between the theta oscillations and the elicited gamma-band activity reflects local spiking activity, which probably occurs for the revision of the predictive model developed in the higher levels of the hierarchy, which is locked to the phase of the slow theta oscillations, during the period in which the excitability of the neural population is higher, hence signaling the time window for updating the model. Further studies to address the information flow between the cortical structures are required to prove this hypothesis.

Investigating the neural response to omission of tones is of specific interest in the framework of predictive coding, since it reflects an elicited response to violation of a sequence without any feedforward propagation of a sensory input (Bekinschtein et al., 2009), and therefore the neural response can be considered to reflect pure prediction (Chennu et al., 2016; SanMiguel et al., 2013). A recent study by Chennu et al. (Chennu et al., 2013) showed that the elicited mismatch response can be best explained when assuming top-down driven inputs in the dynamic causal modeling in higher-order cortical areas. Interestingly, it has been demonstrated that evoked omission responses are sensitive not only to the timing of the stimulus, but also to its predicted identity (Auksztulewicz et al., 2018). Our results on the neural correlates of the omission response do not contradict previous findings related to the omission response. In the previous studies both the deviant and omission stimuli involved manipulation at the same hierarchical level of the stimulus structure (Chennu et al., 2013; Chennu et al., 2016; Wacongne et al., 2011), which made the comparison between the presence and absence of a feedforward input feasible. In our study however, the omission or manipulation of the last chord affected the rhythmic structure of the stimulus, the degree of which was not equal for the two aforementioned conditions. As the omission deviant was delivered in the context of an oddball paradigm and was only different in the absence of the last chord, the elicited MMN reflects a sensory error which was not proceeded by later ERP and oscillatory activities that reflect a model update.

This study addressed the neural oscillatory activity underlying rhythm processing. The results also shed light on the underlying mechanisms of predictive coding in terms of how the predictive error signal is processed and how the internal model is updated when confronting an input that violates the abstracted regularities. Further studies are required to address how this mechanism functions in processing temporal structures through other sensory modalities and the causal interactions in the neural networks that give rise to the observed activity. In addition, it has been shown that newborn infants develop expectation for the onset of rhythmic cycles and create a mismatch response to omission of the downbeat. An interesting question is how the mechanisms involved in the predictive coding of temporal structures, which are widely acknowledged to be an important feature of both music and language, evolve in the course of development and what are the differences between adults and newborns in terms of the mechanisms involved in creating a mismatch response.

**Acknowledgements**

The authors would like to thank Pauline Brunel for recruiting the participants and performing the EEG recordings.

**Funding**

This research was funded by the PhD Eiffel Excellence Scholarship, French Ministry of Foreign Affairs as well as the Iran Cognitive Sciences and Technologies Council (Neurobiom-Iran).

**Declarations of interest: none**

**References**

Akrami, H., Moghimi, S., 2017. Culture modulates the brain response to harmonic violations: an eeg study on hierarchical syntactic structure in music. Frontiers in human neuroscience 11, 591.
Auksztulewicz, R., Schwiedrzik, C.M., Thesen, T., Doyle, W., Devinsky, O., Nobre, A.C., Schroeder, C.E., Friston, K.J., Melloni, L., 2018. Not all predictions are equal:"What" and "when" predictions modulate activity in auditory cortex through different mechanisms. Journal of Neuroscience 38, 8680-8693.
Axmacher, N., Henseler, M.M., Jensen, O., Weinreich, I., Elger, C.E., Fell, J., 2010. Cross-frequency coupling supports multi-item working memory in the human hippocampus. Proceedings of the National Academy of Sciences 107, 3228-3233.
Başar-Eroglu, C., Başar, E., 1991. A compound P300-40Hz response of the cat hippocampus. International Journal of Neuroscience 60, 227-237.
Bastos, A.M., Usrey, W.M., Adams, R.A., Mangun, G.R., Fries, P., Friston, K.J., 2012. Canonical microcircuits for predictive coding. Neuron 76, 695-711.
Bekinschtein, T.A., Dehaene, S., Rohaut, B., Tadel, F., Cohen, L., Naccache, L., 2009. Neural signature of the conscious processing of auditory regularities. Proceedings of the National Academy of Sciences 106, 1672-1677.
Bergmann, T.O., Born, J., 2018. Phase-amplitude coupling: a general mechanism for memory processing and synaptic plasticity? Neuron 97, 10-13.
Bhattacharya, J., Petsche, H., Pereda, E., 2001. Long-range synchrony in the γ band: role in music perception. Journal of Neuroscience 21, 6329-6337.

Bouwer, F.L., Honing, H., 2015. Temporal attending and prediction influence the perception of metrical rhythm: evidence from reaction times and ERPs. Frontiers in psychology 6, 1094.
Bouwer, F.L., Van Zuijen, T.L., Honing, H., 2014. Beat processing is pre-attentive for metrically simple rhythms with clear accents: an ERP study. PloS one 9, e97467.
Bouwer, F.L., Werner, C.M., Knetemann, M., Honing, H., 2016. Disentangling beat perception from sequential learning and examining the influence of attention and musical abilities on ERP responses to rhythm. Neuropsychologia 85, 80-90.
Buzsáki, G., Draguhn, A., 2004. Neuronal oscillations in cortical networks. Science 304, 1926-1929.
Canolty, R.T., Edwards, E., Dalal, S.S., Soltani, M., Nagarajan, S.S., Kirsch, H.E., Berger, M.S., Barbaro, N.M., Knight, R.T., 2006. High gamma power is phase-locked to theta oscillations in human neocortex. Science 313, 1626-1628.
Canolty, R.T., Knight, R.T., 2010. The functional role of cross-frequency coupling. Trends in cognitive sciences 14, 506-515.
Chacko, R.V., Kim, B., Jung, S.W., Daitch, A.L., Roland, J.L., Metcalf, N.V., Corbetta, M., Shulman, G.L., Leuthardt, E.C., 2018. Distinct phase-amplitude couplings distinguish cognitive processes in human attention. NeuroImage 175, 111-121.
Chao, Z.C., Takaura, K., Wang, L., Fujii, N., Dehaene, S., 2018. Large-scale cortical networks for hierarchical prediction and prediction error in the primate brain. Neuron 100, 1252-1266. e1253.
Chennu, S., Noreika, V., Gueorguiev, D., Blenkmann, A., Kochen, S., Ibánez, A., Owen, A.M., Bekinschtein, T.A., 2013. Expectation and attention in hierarchical auditory prediction. Journal of Neuroscience 33, 11194-11205.
Chennu, S., Noreika, V., Gueorguiev, D., Shtyrov, Y., Bekinschtein, T.A., Henson, R., 2016. Silent expectations: dynamic causal modeling of cortical prediction and attention to sounds that weren't. Journal of Neuroscience 36, 8305-8316.
Cheung, V.K., Harrison, P.M., Meyer, L., Pearce, M.T., Haynes, J.-D., Koelsch, S., 2019. Uncertainty and surprise jointly predict musical pleasure and amygdala, hippocampus, and auditory cortex activity. Current Biology 29, 4084-4092. e4084.
Combrisson, E., Perrone-Bertolotti, M., Soto, J.L., Alamian, G., Kahane, P., Lachaux, J.-P., Guillot, A., Jerbi, K., 2017. From intentions to actions: neural oscillations encode motor processes through phase, amplitude and phase-amplitude coupling. NeuroImage 147, 473-487.
Curtis, C.E., D'Esposito, M., 2003. Persistent activity in the prefrontal cortex during working memory. Trends in cognitive sciences 7, 415-423.
Dehaene, S., Meyniel, F., Wacongne, C., Wang, L., Pallier, C., 2015. The neural representation of sequences: from transition probabilities to algebraic patterns and linguistic trees. Neuron 88, 2-19.
Delorme, A., Makeig, S., 2004. EEGLAB: an open source toolbox for analysis of single-trial EEG dynamics including independent component analysis. Journal of neuroscience methods 134, 9-21.
Doeller, C.F., Opitz, B., Mecklinger, A., Krick, C., Reith, W., Schröger, E., 2003. Prefrontal cortex involvement in preattentive auditory deviance detection:: neuroimaging and electrophysiological evidence. NeuroImage 20, 1270-1282.
Dürschmid, S., Edwards, E., Reichert, C., Dewar, C., Hinrichs, H., Heinze, H.-J., Kirsch, H.E., Dalal, S.S., Deouell, L.Y., Knight, R.T., 2016. Hierarchy of prediction errors for auditory events in human temporal and frontal cortex. Proceedings of the National Academy of Sciences 113, 6755-6760.
El Karoui, I., King, J.-R., Sitt, J., Meyniel, F., Van Gaal, S., Hasboun, D., Adam, C., Navarro, V., Baulac, M., Dehaene, S., 2014. Event-related potential, time-frequency, and functional connectivity facets of local and global auditory novelty processing: an intracranial study in humans. Cerebral Cortex 25, 4203-4212.
Elliott, M.T., Wing, A.M., Welchman, A.E., 2014. Moving in time: Bayesian causal inference explains movement coordination to auditory beats. Proceedings of the Royal Society B: Biological Sciences 281, 20140751.
Engel, A.K., Gerloff, C., Hilgetag, C.C., Nolte, G., 2013. Intrinsic coupling modes: multiscale interactions in ongoing brain activity. Neuron 80, 867-886.

Fell, J., Axmacher, N., 2011. The role of phase synchronization in memory processes. Nature reviews neuroscience 12, 105.
Friedman, D., Cycowicz, Y.M., Gaeta, H., 2001. The novelty P3: an event-related brain potential (ERP) sign of the brain's evaluation of novelty. Neuroscience & Biobehavioral Reviews 25, 355-373.
Friston, K., 2002. Beyond phrenology: what can neuroimaging tell us about distributed circuitry? Annual review of neuroscience 25, 221-250.
Friston, K., 2005. A theory of cortical responses. Philosophical transactions of the Royal Society B: Biological sciences 360, 815-836.
Friston, K., 2010. The free-energy principle: a unified brain theory? Nature reviews neuroscience 11, 127-138.
Garrido, M.I., Friston, K.J., Kiebel, S.J., Stephan, K.E., Baldeweg, T., Kilner, J.M., 2008. The functional anatomy of the MMN: a DCM study of the roving paradigm. NeuroImage 42, 936-944.
Garrido, M.I., Kilner, J.M., Kiebel, S.J., Friston, K.J., 2007. Evoked brain responses are generated by feedback loops. Proceedings of the National Academy of Sciences 104, 20961-20966.
Garrido, M.I., Kilner, J.M., Kiebel, S.J., Friston, K.J., 2009. Dynamic causal modeling of the response to frequency deviants. Journal of Neurophysiology 101, 2620-2631.
Geiser, E., Ziegler, E., Jancke, L., Meyer, M., 2009. Early electrophysiological correlates of meter and rhythm processing in music perception. cortex 45, 93-102.
Gilbert, P.E., Kesner, R.P., 2006. The role of the dorsal CA3 hippocampal subregion in spatial working memory and pattern separation. Behavioural brain research 169, 142-149.
Gold, B.P., Pearce, M.T., Mas-Herrero, E., Dagher, A., Zatorre, R.J., 2019. Predictability and uncertainty in the pleasure of music: a reward for learning? Journal of Neuroscience 39, 9397-9409.
Grahn, J.A., 2012. Neural mechanisms of rhythm perception: current findings and future perspectives. Topics in cognitive science 4, 585-606.
Haegens, S., Nácher, V., Luna, R., Romo, R., Jensen, O., 2011. α-Oscillations in the monkey sensorimotor network influence discrimination performance by rhythmical inhibition of neuronal spiking. Proceedings of the National Academy of Sciences 108, 19377-19382.
Hansen, N.C., Pearce, M.T., 2014. Predictive uncertainty in auditory sequence processing. Frontiers in psychology 5, 1052.
Hashemi, N.S., Dehnavi, F., Moghimi, S., Ghorbani, M., 2019. Slow spindles are associated with cortical high frequency activity. NeuroImage 189, 71-84.
Haumann, N.T., Vuust, P., Bertelsen, F., Garza-Villarreal, E.A., 2018. Influence of musical enculturation on brain responses to metric deviants. Frontiers in neuroscience 12, 218.
Helfrich, R.F., Mander, B.A., Jagust, W.J., Knight, R.T., Walker, M.P., 2018. Old brains come uncoupled in sleep: slow wave-spindle synchrony, brain atrophy, and forgetting. Neuron 97, 221-230. e224.
Herrmann, C.S., Munk, M.H., Engel, A.K., 2004. Cognitive functions of gamma-band activity: memory match and utilization. Trends in cognitive sciences 8, 347-355.
Honing, H., Ladinig, O., Háden, G.P., Winkler, I., 2009. Is beat induction innate or learned? Probing emergent meter perception in adults and newborns using event-related brain potentials. Annals of the New York Academy of Sciences 1169, 93-96.
Hyafil, A., Giraud, A.-L., Fontolan, L., Gutkin, B., 2015. Neural cross-frequency coupling: connecting architectures, mechanisms, and functions. Trends in neurosciences 38, 725-740.
James, C.E., Michel, C.M., Britz, J., Vuilleumier, P., Hauert, C.A., 2012. Rhythm evokes action: Early processing of metric deviances in expressive music by experts and laymen revealed by ERP source imaging. Human brain mapping 33, 2751-2767.
Janata, P., 1995. ERP measures assay the degree of expectancy violation of harmonic contexts in music. Journal of Cognitive neuroscience 7, 153-164.
Jongsma, M.L., Desain, P., Honing, H., 2004. Rhythmic context influences the auditory evoked potentials of musicians and nonmusicians. Biological psychology 66, 129-152.
Kaiser, J., Hertrich, I., Ackermann, H., Mathiak, K., Lutzenberger, W., 2005. Hearing lips: gamma-band activity during audiovisual speech perception. Cerebral Cortex 15, 646-653.

Kaiser, J., Leiberg, S., Rust, H., Lutzenberger, W., 2007. Prefrontal gamma-band activity distinguishes between sound durations. Brain research 1139, 153-162.
Kaiser, J., Lutzenberger, W., Ackermann, H., Birbaumer, N., 2002. Dynamics of gamma-band activity induced by auditory pattern changes in humans. Cerebral Cortex 12, 212-221.
Kaiser, J., Ripper, B., Birbaumer, N., Lutzenberger, W., 2003. Dynamics of gamma-band activity in human magnetoencephalogram during auditory pattern working memory. NeuroImage 20, 816-827.
Kanai, R., Komura, Y., Shipp, S., Friston, K., 2015. Cerebral hierarchies: predictive processing, precision and the pulvinar. Philosophical transactions of the Royal Society B: Biological sciences 370, 20140169.
Koelsch, S., Vuust, P., Friston, K., 2019. Predictive processes and the peculiar case of music. Trends in cognitive sciences 23, 63-77.
Ladinig, O., Honing, H., Hááden, G., Winkler, I., 2009. Probing attentive and preattentive emergent meter in adult listeners without extensive music training. Music Perception: An Interdisciplinary Journal 26, 377-386.
Lakatos, P., Shah, A.S., Knuth, K.H., Ulbert, I., Karmos, G., Schroeder, C.E., 2005. An oscillatory hierarchy controlling neuronal excitability and stimulus processing in the auditory cortex. Journal of Neurophysiology 94, 1904-1911.
Lappe, C., Lappe, M., Pantev, C., 2016. Differential processing of melodic, rhythmic and simple tone deviations in musicians-an MEG study. NeuroImage 124, 898-905.
Lappe, C., Steinsträter, O., Pantev, C., 2013. Rhythmic and melodic deviations in musical sequences recruit different cortical areas for mismatch detection. Frontiers in human neuroscience 7, 260.
Lappe, C., Trainor, L.J., Herholz, S.C., Pantev, C., 2011. Cortical plasticity induced by short-term multimodal musical rhythm training. PloS one 6, e21493.
Lelo-de-Larrea-Mancera, E.S., Rodríguez-Agudelo, Y., Solís-Vivanco, R., 2017. Musical rhythm and pitch: A differential effect on auditory dynamics as revealed by the N1/MMN/P3a complex. Neuropsychologia 100, 44-50.
Longuet-Higgins, H.C., Lee, C.S., 1984. The rhythmic interpretation of monophonic music. Music Perception: An Interdisciplinary Journal 1, 424-441.
Lumaca, M., Trusbak Haumann, N., Brattico, E., Grube, M., Vuust, P., 2018. Weighting of neural prediction error by rhythmic complexity: a predictive coding account using Mismatch Negativity. European Journal of Neuroscience.
Maris, E., Oostenveld, R., 2007. Nonparametric statistical testing of EEG-and MEG-data. Journal of neuroscience methods 164, 177-190.
Molholm, S., Martinez, A., Ritter, W., Javitt, D.C., Foxe, J.J., 2005. The neural circuitry of pre-attentive auditory change-detection: an fMRI study of pitch and duration mismatch negativity generators. Cerebral Cortex 15, 545-551.
Noesselt, T., Shah, N.J., Jäncke, L., 2003. Top-down and bottom-up modulation of language related areas–an fMRI study. BMC neuroscience 4, 13.
Oostenveld, R., Fries, P., Maris, E., Schoffelen, J.-M., 2011. FieldTrip: open source software for advanced analysis of MEG, EEG, and invasive electrophysiological data. Computational intelligence and neuroscience 2011, 1.
Park, H.-J., Friston, K., 2013. Structural and functional brain networks: from connections to cognition. Science 342, 1238411.
Patel, A.D., Daniele, J.R., 2003. An empirical comparison of rhythm in language and music. Cognition 87, B35-B45.
Pearce, M.T., Wiggins, G.A., 2012. Auditory expectation: the information dynamics of music perception and cognition. Topics in cognitive science 4, 625-652.
Phillips, H.N., Blenkmann, A., Hughes, L.E., Bekinschtein, T.A., Rowe, J.B., 2015. Hierarchical organization of frontotemporal networks for the prediction of stimuli across multiple dimensions. Journal of Neuroscience 35, 9255-9264.
Rasch, B., Born, J., 2013. About sleep's role in memory. Physiological reviews 93, 681-766.

Recasens, M., Gross, J., Uhlhaas, P.J., 2018. Low-frequency oscillatory correlates of auditory predictive processing in cortical-subcortical networks: A MEG-Study. Scientific reports 8, 14007.
Rinne, T., Degerman, A., Alho, K., 2005. Superior temporal and inferior frontal cortices are activated by infrequent sound duration decrements: an fMRI study. NeuroImage 26, 66-72.
Rohrmeier, M.A., Koelsch, S., 2012. Predictive information processing in music cognition. A critical review. International Journal of Psychophysiology 83, 164-175.
Salimpour, Y., Anderson, W.S., 2019. Cross-Frequency Coupling Based Neuromodulation for Treating Neurological Disorders. Frontiers in neuroscience 13.
SanMiguel, I., Widmann, A., Bendixen, A., Trujillo-Barreto, N., Schröger, E., 2013. Hearing silences: human auditory processing relies on preactivation of sound-specific brain activity patterns. Journal of Neuroscience 33, 8633-8639.
Snyder, J.S., Large, E.W., 2005. Gamma-band activity reflects the metric structure of rhythmic tone sequences. Cognitive brain research 24, 117-126.
Sokolov, A., Pavlova, M., Lutzenberger, W., Birbaumer, N., 2004. Reciprocal modulation of neuromagnetic induced gamma activity by attention in the human visual and auditory cortex. NeuroImage 22, 521-529.
Staresina, B.P., Bergmann, T.O., Bonnefond, M., Van Der Meij, R., Jensen, O., Deuker, L., Elger, C.E., Axmacher, N., Fell, J., 2015. Hierarchical nesting of slow oscillations, spindles and ripples in the human hippocampus during sleep. Nature neuroscience 18, 1679.
Summerfield, C., Egner, T., Greene, M., Koechlin, E., Mangels, J., Hirsch, J., 2006. Predictive codes for forthcoming perception in the frontal cortex. Science 314, 1311-1314.
Tallon-Baudry, C., Bertrand, O., 1999. Oscillatory gamma activity in humans and its role in object representation. Trends in cognitive sciences 3, 151-162.
Tort, A.B., Komorowski, R., Eichenbaum, H., Kopell, N., 2010. Measuring phase-amplitude coupling between neuronal oscillations of different frequencies. Journal of Neurophysiology 104, 1195-1210.
Tort, A.B., Komorowski, R.W., Manns, J.R., Kopell, N.J., Eichenbaum, H., 2009. Theta–gamma coupling increases during the learning of item–context associations. Proceedings of the National Academy of Sciences 106, 20942-20947.
Trainor, L.J., McDonald, K.L., Alain, C., 2002. Automatic and controlled processing of melodic contour and interval information measured by electrical brain activity. Journal of Cognitive neuroscience 14, 430-442.
Uhrig, L., Dehaene, S., Jarraya, B., 2014. A hierarchy of responses to auditory regularities in the macaque brain. Journal of Neuroscience 34, 1127-1132.
Vuust, P., Liikala, L., Näätänen, R., Brattico, P., Brattico, E., 2016. Comprehensive auditory discrimination profiles recorded with a fast parametric musical multi-feature mismatch negativity paradigm. Clinical Neurophysiology 127, 2065-2077.
Vuust, P., Ostergaard, L., Pallesen, K.J., Bailey, C., Roepstorff, A., 2009. Predictive coding of music–brain responses to rhythmic incongruity. cortex 45, 80-92.
Vuust, P., Pallesen, K.J., Bailey, C., Van Zuijen, T.L., Gjedde, A., Roepstorff, A., Østergaard, L., 2005. To musicians, the message is in the meter: pre-attentive neuronal responses to incongruent rhythm are left-lateralized in musicians. NeuroImage 24, 560-564.
Wacongne, C., Labyt, E., van Wassenhove, V., Bekinschtein, T., Naccache, L., Dehaene, S., 2011. Evidence for a hierarchy of predictions and prediction errors in human cortex. Proceedings of the National Academy of Sciences 108, 20754-20759.
Wang, L., Uhrig, L., Jarraya, B., Dehaene, S., 2015. Representation of numerical and sequential patterns in macaque and human brains. Current Biology 25, 1966-1974.
Zanto, T.P., Large, E.W., Fuchs, A., Kelso, J.S., 2005. Gamma-band responses to perturbed auditory sequences: evidence for synchronization of perceptual processes. Music Perception: An Interdisciplinary Journal 22, 531-547.
Zhao, T.C., Lam, H.G., Sohi, H., Kuhl, P.K., 2017. Neural processing of musical meter in musicians and non-musicians. Neuropsychologia 106, 289-297.

# Supplementary Information

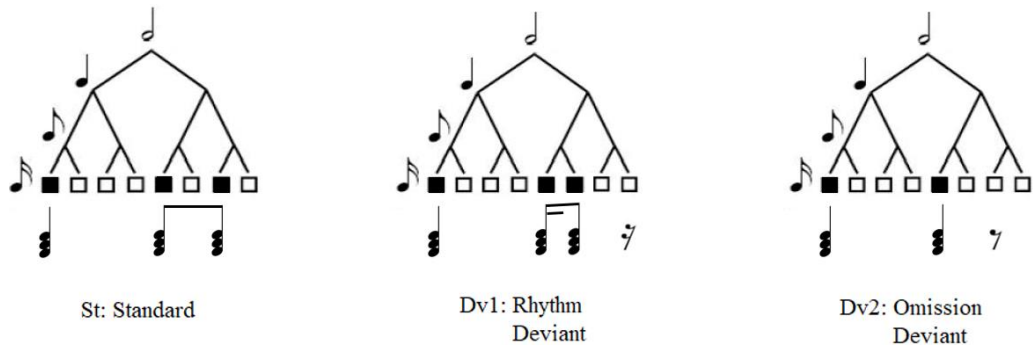

**Figure S1. Rhythm tree and the corresponding events for the three standard (St), rhythm deviant (Dv1), and omission deviant (Dv2) conditions.**

## Control study to control the effect of stimulus structure on the omission response

### Participants

Ten right-handed healthy volunteers other than those who participated in the main experiment, with inclusion criteria similar to that described in the manuscript, participated in this follow-up study.

### Experiment design

For this experiment, we used four stimuli: the standard stimulus, the rhythm deviant stimulus, the omission stimulus (omission I), and an additional omission stimulus in which the last chord of the rhythm deviant was silenced (omission II).

The stimuli were delivered in the context of an oddball paradigm. The experimental session consisted of three blocks. The first included high-probability standard stimuli (p = 0.85, 595 trials) interspersed with the omission I deviant (p = 0.15, 105 trials). The second block included high-probability rhythm deviant stimuli as standard stimuli (p = 0.85, 595 trials) interspersed with the omission II deviant (p = 0.85, 595 trials). The third block was the same as the first, except that the rhythm deviant was used as the deviant stimulus. The order of the deviant stimuli in each block was pseudo-randomized among the standard trials, enforcing three to seven standard stimuli between successive deviant trials. These blocks were presented randomly. Stimuli were delivered through two custom made speakers at 65 dB SPL using Psychtoolbox MATLAB. Participants sat in a comfortable chair in dim light and were instructed to watch a silent movie (March of the Penguins, Warner, ASIN B000BI5KV0). The total duration of the experiment was ~42 min.

**EEG acquisition and preprocessing and ERP analysis**

All the EEG acquisition and preprocessing steps were the same as those described in the main manuscript. In addition, the same procedure was applied to create the ERP results and perform the statistical analyses.

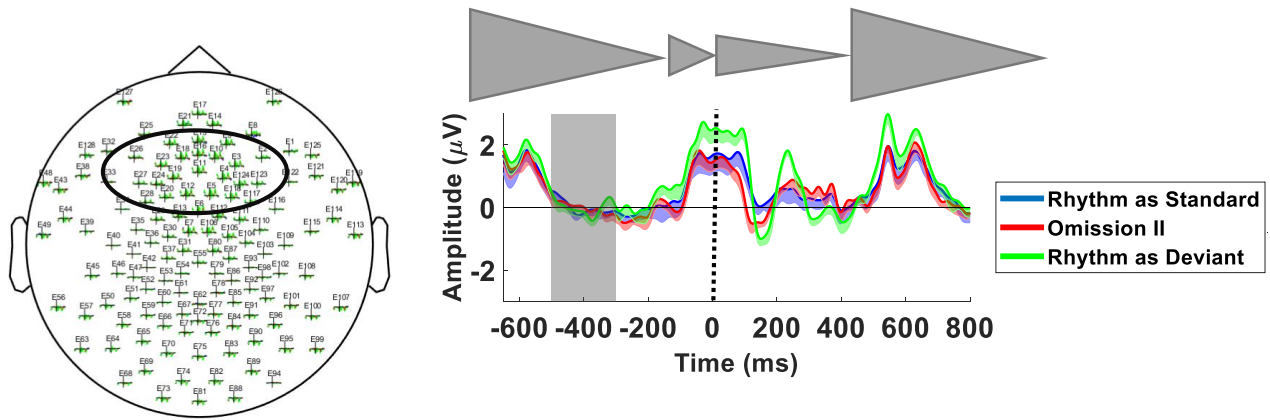

Figure S2. Event-locked analysis of rhythm and omission II conditions. The onset of the deviant chord was set to zero and the next trial started at 450 ms. For all conditions, the baseline was set to 250 to 450 ms from the onset of the first chord. Cluster-based statistics revealed four spatiotemporal clusters for the rhythm deviant condition: (1) a negative cluster (p = 0.002, corrected), comprising frontal and frontocentral electrodes and extending over 133 to 190 ms post-final chord, (2) a posterior positive cluster (p = 0.02, corrected), synchronous with the first cluster, 129 to 186 ms post-final chord, (3) a negative posterior cluster (p = 0.014, corrected), extending over 223 to 258 ms post-final chord, and (4) a positive frontal cluster (p = 0.034, corrected), synchronous with the third cluster, 209 to 254 ms post-final chord. Thus, in the follow-up control experiment, the rhythm deviant condition again elicited an MMN response followed by a P3a component. However, cluster-based statistics revealed only two spatiotemporal clusters for the omission II condition: (1) a negative cluster (p = 0.008, corrected), comprising frontal and frontocentral electrodes and extending over 143 to 174 ms post-final chord, and (2) a posterior positive cluster (p = 0.002, corrected), synchronous with the first cluster, 136 to 168 ms post-final chord. The timing of this cluster matched that of the MMN and no cluster in the timing window of P3a was verified to be significant. The MMN corresponding to the rhythm deviant was significantly larger than that corresponding to the omission II condition (t=2.65, p = 0.0328).

**Additional figures**

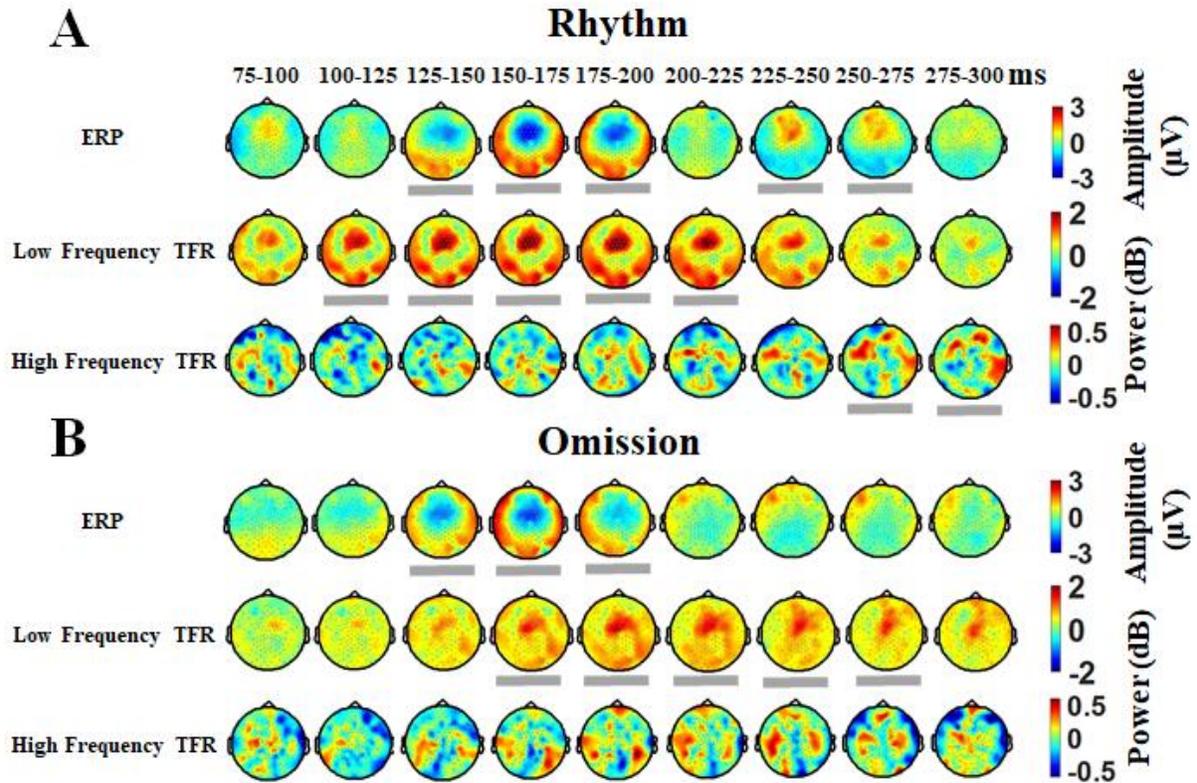

Figure S3. Topographical distribution of the clusters. Columns correspond to each time window from 75 to 300 ms and rows correspond to the ERP, low frequency TFR, and high frequency TFR, respectively. (A) Topographical distribution of the clusters corresponding to the rhythm deviant. (B) Topographical distribution of the clusters corresponding to the omission deviant.

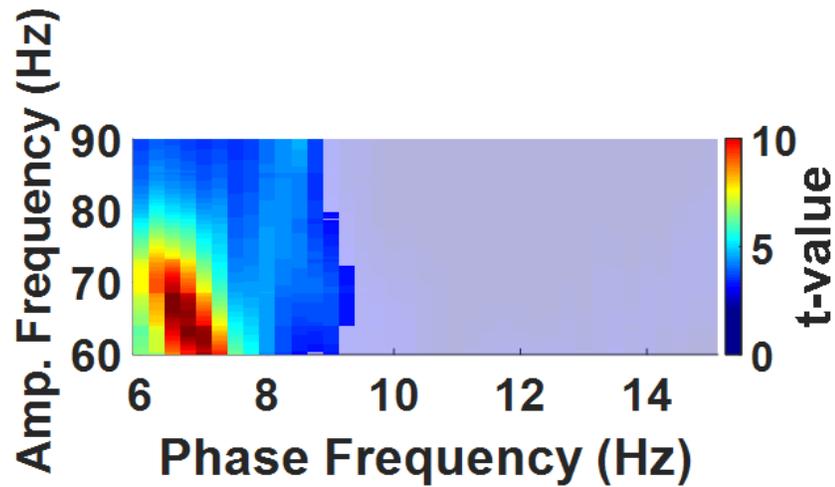

**Figure S4. Cluster-based permutation results on phase-amplitude coupling over the 450-ms window, corresponding to the rhythm deviant condition and epoch-shuffled surrogate data – the same data as that used for the original PAC analysis. The grey regions correspond to frequency pairs for which the permutation analysis did not show significant PAC.**